\documentclass[a4paper, times]{article}
\usepackage[margin=1in]{geometry}
\usepackage{amsmath}
\usepackage{amsfonts}
\usepackage{amssymb}
\usepackage{mathtools}

\usepackage[colorlinks,bookmarksopen,bookmarksnumbered,citecolor=red,urlcolor=red]{hyperref}

\begin{document}

\title{Further investigation of the spurious interface fragmentation in multiphase Smoothed Particle Hydrodynamics}

\author{K.~Szewc, M.T.~Lewandowski\\
  Institute of Fluid-Flow Machinery, Polish Academy of Sciences,\\ 
  ul. Fiszera 14, 80231 Gda\'nsk, Poland\\
  \texttt{kszewc@imp.gda.pl}}

\maketitle

\begin{abstract}

This article presents results of further investigation of the problem of spurious interface fragmentation in the multiphase SPH.
In order to remove arising instabilities, many authors introduced the 
artificial interface correction procedure.
In the present paper we show that the interface instabilities are physical and the introduction of the interface correction procedure may leads to non-physical solutions. We also explain the puzzling relation between the parameters $\varepsilon$ and $h$. The analysis is performed on the basis of the stability analysis and numerical calculations using SPH and Volume Of Fluid (VOF) approach.

\end{abstract}

\vspace{-6pt}

\section{Introduction}

In the articles \cite{Colagrossi and Landrini 2003}, \cite{Das and Das 2009}, \cite{Grenier et al. 2009}, and in details \cite{Szewc et al. 2015}
the authors discuss the issue of sub-kernel spurious interface fragmentation occurring in SPH applied for multiphase flows.
One of the reasons for this adverse phenomenon is a lack of mechanisms assuring immisicibility of phases. 
In order to reduce the negative effects of this issue, the authors proposed different interface correction terms (added to the Navier-Stokes equation).
For example, in \cite{Szewc et al. 2015} the authors proposed the correction term in the form
\begin{equation} \label{xi}
\Xi_a = \frac{\varepsilon}{m_a} \sum_{\substack{b \\ c_b \neq c_a}} \left( \frac{1}{\Theta_a} + \frac{1}{\Theta_b} \right) \nabla_a W_{ab}(h),
\end{equation}
where $\varepsilon$ is the empirical parameter, $m$ is particle mass, $W_{ab}(h)$ is the kernel function, $h$ is smoothing length (scale of smoothing), while $\Theta_a = \sum_b W_{ab}(h)$. Variable $c$ is the color function used to indicate phases.
Based on these numerical experiments the authors demonstrated the dependence of the minimal value of $\varepsilon$ which gives the stable interface on the kernel smoothing length $h$
\begin{equation} \label{e-h}
  \varepsilon_{min} \sim \frac{1}{h}.
\end{equation}
The relation (\ref{e-h}) was proposed on the basis of the bubble rising in water test case with no surface tension. This test case involves a single bubble of the radius $R$ placed at
\begin{equation}
  \left(x - \frac{L_x}{2}\right)^2 + \left(y - \frac{L_y}{2}\right)^2 <  R^2,
\end{equation}
where $L_x=L=6R$ and $L_y=12R$ are the domain edge sizes.
The density and viscosity ratios between liquid and gas phases are respectively $128$ and $1000$. 

This case was first calculated by \emph{Sussman et al. (1994)}~\cite{Sussman et al. 1994} using the level-set approach. These authors presented the solution corresponding to:
\begin{equation}
  Re = \frac{\sqrt{8R^3g}}{\nu}=1000, \quad Bo=\frac{4\varrho g R^2}{\sigma}=200,
\end{equation}
where $Re$ is the Reynolds number, $Bo$ is the Bond number, $g$ is the gravitational acceleration, $\sigma$ is the surface tension coefficient, while $\varrho$ and $\nu$ are respectively the density and the viscosity of the liquid phase. It is important here to note, that in the numerical experiment of \emph{Sussman et al.} the surface tension is present.

The first SPH calculation of this test case was performed by \emph{Colagrossi and Landrini (2003)}~\cite{Colagrossi and Landrini 2003} with the newly proposed multi-phase SPH formulation. The authors noted that $Bo=200$ corresponds to very small surface tension compared with other effects, so they decided that the surface tension effects are negligible. As a result authors could not obtain stable solution, and, therefore, they proposed an additional artificial term to the Navier-Stokes equation responsible for the interface stabilization
\begin{equation} \label{colagrossi-landrini}
  \Xi_a = -\varepsilon \sum_{\substack{b \\ c_b \neq c_a}} \frac{m_b}{\varrho_b} (\varrho^2_a + \varrho_b^2)\nabla_a W_{ab}(h).
\end{equation}
A few years later, \emph{Grenier et al. (2009)}~\cite{Grenier et al. 2009} proposed a novel SPH multi-phase formulation. Since the authors did not considered surface-tension effects in their work, they decided to perform calculations of this test case with no surface-tension. As previously, in order to obtain stable bubbles, the authors introduced an additional term to the N-S equation in a form: 
\begin{equation} \label{grenier sharpness}
  \Xi_a = \varepsilon \sum_{\substack{b \\ c_b \neq c_a}} \frac{m_b}{\varrho_b} \left( \left| \frac{p_a}{\Gamma_a} \right| + \left| \frac{p_b}{\Gamma_b} \right| \right) \nabla_a W_{ab}(h),
  \quad \Gamma_a = \sum_b \frac{m_b}{\varrho_b} W_{ab}(h).
\end{equation}
Later, \emph{Szewc et al. (2015)}\cite{Szewc et al. 2015} presented the detailed analysis of the influence of the correction in the form of Eq.~(\ref{xi}). As like predecessors, their analysis is based on the same test case with no surface tension.

The main aim of this paper is to show that the instabilities obtained in the bubble rising in liquid test case proposed by \emph{Sussman et al. (1994)} when the surface tension is neglegible are physically correct. We show that the introduction of the interface correction term proposed in \cite{Szewc et al. 2015}, \cite{Colagrossi and Landrini 2003}, and \cite{Grenier et al. 2009} may lead to non-physical solutions.

The paper is organized as follows: in Section~\ref{sec:K-H instability} the Kevin-Helmholtz stability analysis is presented to show that in the introduced test case the bubble interface is unstable when no surface tension is present. Then, in Section~\ref{sec:surface tension} we show the stabilization effects of surface tension on the interface. In that section, we also discuss a role of the interface correction and its influence on the SPH solutions. Finally, in Section~\ref{sec:e-h}, based on the stability analysis, we explain the relation between $\varepsilon_{min}$ and $h$, Eq.~(\ref{e-h}), obtained in~\cite{Szewc et al. 2015}.

In the paper we use the `numerical units', in which all the variables are non-dimentionalized with $R = 1$ cm, $g = 1$ cm/s$^{2}$ and $\varrho=1$ g/cm$^{3}$. The stability analysis is based on the course on hydrodynamic stability by \emph{A.J.~Mestel} (Imperial Collage, UK)~\cite{Mestel}.

\section{The Kelvin-Helmholtz instability analysis}\label{sec:K-H instability}
\begin{figure*}
  \centering
  \includegraphics[width=0.55\textwidth]{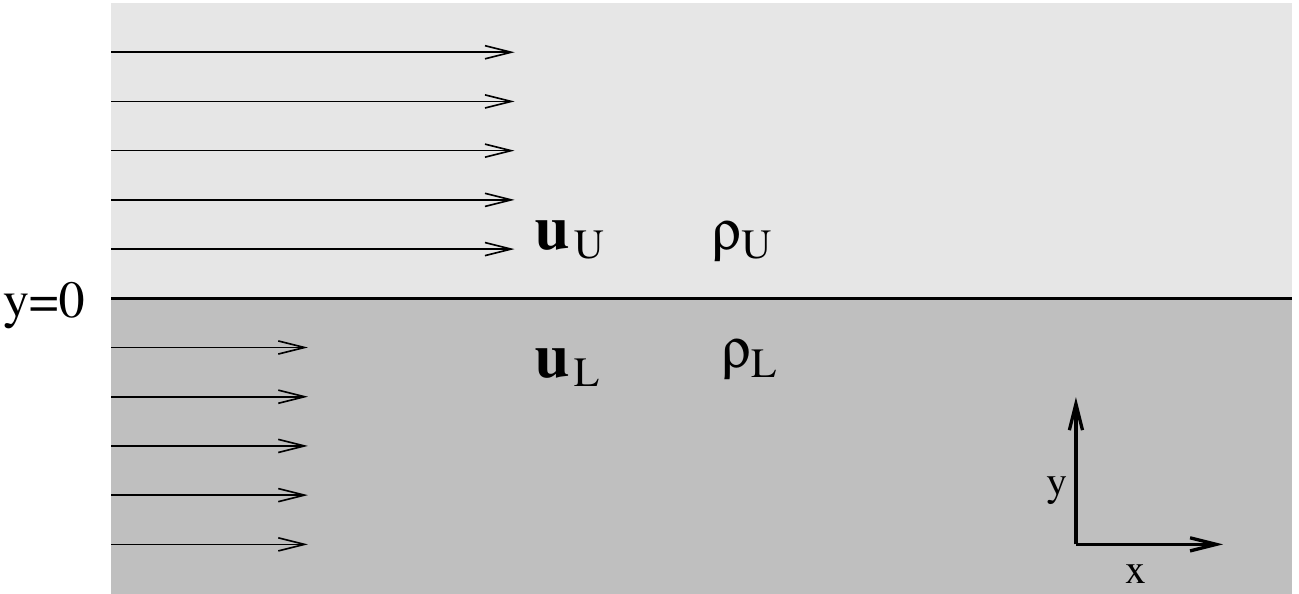}
  \caption{Initial configuration for the stability analysis.}
  \label{fig:initial-configuration}
\end{figure*}

\begin{figure*}
  \centering
  \includegraphics[width=0.3\textwidth]{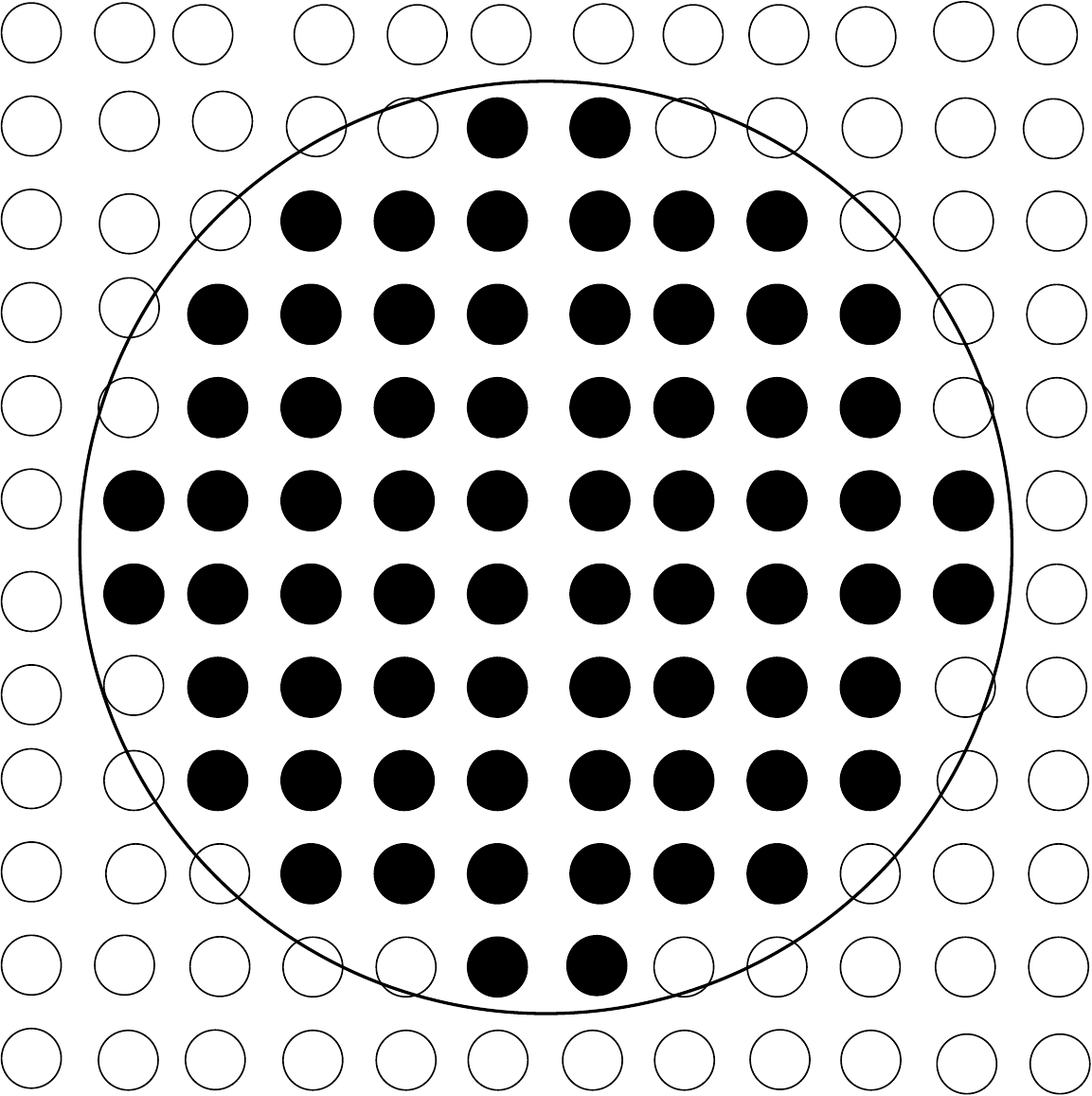}
  \caption{Local disturbance of the interface shape due to the discretization.}
  \label{fig:disturbance}
\end{figure*}

Let us consider two fluid regions: upper ($y>0$) with the initially uniform velocity $\mathbf u_u=(u_U, 0)$ and the density $\varrho_U$ and lower ($y<0$) with the velocity $\mathbf u_L = (u_L, 0)$ and the density $\varrho_L$, see Fig.~\ref{fig:initial-configuration}.
The system is under the gravitational acceleration $\mathbf g=(0,-g)$. Both fluid are assumed to be inviscid. The interface shape can be described by equation
\begin{equation}
  y = \epsilon \xi(x,t).
\end{equation}
Since, initially, there is no vorticity in the system, the flow is irrotational
\begin{equation}
  \mathbf u = \nabla \varphi, \quad
  \nabla^2 \varphi = 0.
\end{equation}
Then, the variable $\varphi$ can take the form
\begin{equation}
  \varphi =  \left \{
 \begin{array}{cl}
  u_Ux + \epsilon \varphi_U, &  y>0, \\
  u_Lx + \epsilon \varphi_L, &  y<0.  \\
  \end{array}
  \right.
\end{equation}
The kinetic boundary condition at the interface take the form
\begin{equation}
  \frac{d}{dt}(y-\epsilon \xi) = \frac{\partial \varphi}{\partial x} - \epsilon \frac{\partial \xi}{\partial t} - \epsilon \frac{\partial \varphi}{\partial x}\frac{\partial \xi}{\partial x} = 0.
\end{equation}
Neglecting higher order terms of $\epsilon$, we can get the pair of equations for each phase:
\begin{equation} \label{kinetic constraint}
  \frac{\partial \varphi_U}{\partial y} = \frac{\partial \xi}{\partial t} + u_U \frac{\partial \xi}{\partial t}, \quad
  \frac{\partial \varphi_L}{\partial y} = \frac{\partial \xi}{\partial t} + u_L \frac{\partial \xi}{\partial t}.
\end{equation}
The second boundary condition at the interface is the continuity of pressure, which can be satisfied by the time-dependent Bernoulli condition
\begin{equation} \label{Bernoulli}
  p + \left( \varrho \frac{\partial \varphi}{\partial t} + \frac{1}{2}(\nabla \varphi)^2 + gy \right) = const.
\end{equation}
Assuming no presence of surface tension, we can write $p_U=p_L$ at $y=\epsilon \xi$. Neglecting higher terms than first order, we obtain the relation
\begin{equation} \label{pressure constraint}
  \varrho_U\left( \frac{\partial \varphi_U}{\partial t} + u_U\frac{\partial \varphi_U}{\partial x} + g \xi \right) = \varrho_L \left(\frac{\partial \varphi_L}{\partial t} + u_L \frac{\partial \varphi_L}{\partial x} + g \xi \right).
\end{equation}
To leading order, Eqs.~(\ref{kinetic constraint}) and (\ref{pressure constraint}) can be evaluated on $y=0$ rather than $y=\epsilon \xi$.
In the next step, we consider a small disturbance in the form of the plane-wave solution:
\begin{equation} \label{disturbance}
 \xi = \xi_0 e^{ikx+st}, \quad  \varphi_U = \Psi_U(y)e^{ikx+st}, \quad \Psi_L(y)e^{ikx+st}.
\end{equation}
Now, the quantities $\Psi_U(y)$ and $\Psi_L(y)$ should satisfy the following ODEs and boundary conditions:
\begin{equation}
  \frac{d^2 \Psi_U}{dy^2} - k^2\Psi_U = 0,\quad \Psi_U(\infty) = 0,\quad \Psi_U(0) = \xi_0(s+iku_U),
\end{equation}
and
\begin{equation}
  \frac{d^2 \Psi_L}{dy^2} - k^2\Psi_L = 0,\quad \Psi_L(-\infty) = 0,\quad \Psi_L(0) = \xi_0(s+iku_L).
\end{equation}
It gives us
\begin{equation} \label{psi}
  \Psi_U(y) = -\frac{\xi_0}{k}(s+iku_U)e^{-ky}, \quad \Psi_L(y) = \frac{\xi_0}{k}(s+iku_L)e^{ky}.
\end{equation}
The combination of the pressure constraint (\ref{pressure constraint}) with (\ref{disturbance}) and (\ref{psi}) gives the quadratic equation for variable $s$
\begin{equation}
 \varrho_U\left[gk-(s+iku_U)^2\right] =  \varrho_L\left[gk+(s+iku_L)^2\right].
\end{equation}
Solving this equation gives us the dispersion relation in the form
\begin{equation} \label{dispersion relation}
  s = -ik \frac{\varrho_Uu_U + \varrho_Lu_L}{\varrho_U + \varrho_L} \pm \left[\frac{k^2 \varrho_U \varrho_L (u_U-u_L)^2}{(\varrho_U+\varrho_L)^2} + kg \frac{\varrho_U - \varrho_L}{\varrho_U + \varrho_L} \right]^\frac{1}{2}.
\end{equation} 
In order to obtain the non-stable solution the variable $s$ must have a positive real part. It occurs only if 
\begin{equation}
 k^2\varrho_U \varrho_L(u_U-u_L)^2 > kg(\varrho_L^2 - \varrho_U^2).
\end{equation}
Now, let us assume that the described configuration corresponds to the upper part of the bubble considered as a test case: $\varrho_U=1.0$, $\varrho_L=0.001$, $g=1$ (numerical units). 
It is important to note that the physical fields as well as the interface shape are seen by SPH as smoothed in range of $4h$ (compact support Wendland kernel, see \cite{Szewc et al. 2015}). Let us assume the interface disturbance of the wave number $k=2\pi/4h$, originating from smoothing the initial particle distribution, see Fig.~\ref{fig:disturbance}. 
The typical velocity jump $u_U-u_L$, across the interface at beginning of the bubble rise is about $0.2$. It gives us the instability condition
\begin{equation}
  h > 10^{-5}L,
\end{equation}
while, in our test case, we have $h = 0.015625L$. Therefore, due to the Kelvin-Helmholtz instability we are not able to obtain any stable solution in a reasonable resolution when the surface tension effects are omitted.

\section{The interface stability analysis and surface tension} \label{sec:surface tension}
In the previous section we have shown that the degenerated shapes of the interface that we observe in the described bubble rising in liquid test case is the manifestation of the Kelvin-Helmholtz instability. It is commonly known that this type of instabilities can be stabilized by the capillary forces. In this section we apply the stability analysis to show that indeed we can avoid unstable solutions by applying the surface tension force in SPH. In addition, we derive the relation between the surface tension coefficient $\sigma$ and the smoothing length $h$.
The curvature of two-dimensional interface is
\begin{equation}
  \kappa = \nabla \cdot  \widehat{ \mathbf n} = \nabla \cdot \left[ \left(-\epsilon i k \xi_0, 1 \right)e^{ikx + st} \right] = \epsilon k^2 \xi_0 e^{ikx + st}.
\end{equation}
The pressure boundary condition at the interface, cf. Eq.~(\ref{pressure constraint}), when the surface tension is present is replaced by the normal stress condition in the form
\begin{equation} \label{pressure jump with surface tension}
 p_U - p_L = \sigma \kappa.
\end{equation}
Now, the dispersion relation (\ref{dispersion relation}) takes the form
\begin{equation}
  s = -ik \frac{\varrho_U u_U + \varrho_L u_L}{\varrho_U + \varrho_L} \pm \left[ \frac{k^2\varrho_U \varrho_L (u_U - u_L)^2}{(\varrho_U + \varrho_L)^2} + kg\frac{\varrho_U - \varrho_L}{\varrho_U + \varrho_L} - \frac{k^3\sigma}{\varrho_U + \varrho_L} \right]^{\frac{1}{2}}.
\end{equation}
This system is stable if
\begin{equation} \label{f}
  f(k) = \sigma k^3 - \frac{\varrho_U \varrho_L (u_U - u_L)^2}{\varrho_U + \varrho_L} k^2 - (\varrho_U - \varrho_L)gk > 0.
\end{equation}

In order to check whether these prediction agree with the numerical experiment, we decided to perform the simulations of bubble rising in liquid for different values of surface tension coefficient: $\sigma=0.08$, $0.04$, $0.02$, $0.01$, $0.005$, $0.00125$ and three different resolutions: $h=0.046875$, $0.09375$ and $0.1875$ ($h/\Delta r = 2$). The interface sharpness correction was not present. The wavelength number of the disturbance was $k=2\pi/4h=33.5$ for the highest considered resolution and $k=8.375$ for the lowest resolution. The minimal values of $k$ for which according to (\ref{f}), the system is stable, compared with the numerical experiment are presented in Table 1. The obtained numerical results show good agreement with predictions (taking into account all the approximations in the stability analysis). 
\begin{figure}
\caption{The minimal values of $k$ for which system is stable (from stability analysis) compared with the numerical experiments using the SPH method.}
\begin{tabular}{ |c|c|c|c|c| }
  \hline
  \multicolumn{2}{|c|}{}  &\multicolumn{3}{|c|}{Is stable in SPH?} \\
  \hline
   &  & $h=0.046874$ & $h=0.09375$ & $h=0.1875$ \\
   $\sigma$ & $k_{min}$ & $k=33.5$ & $k=16.75$ & $k=8.375$ \\ \hline
   $0.08$ & $3.5$ & yes & yes & yes \\ \hline
   $0.04$ & $5.0$ & yes & yes & no \\ \hline
   $0.02$ & $7.0$ & yes & yes & no \\ \hline
   $0.01$ & $10.0$ & yes & no & no \\ \hline
   $0.005$ & $14.1$ & yes & no & no \\ \hline
   $0.0025$ & $20.0$ & no & no & no \\ \hline
   $0.00125$ & $28.3$ & no & no & no \\ \hline
  
  \hline
\end{tabular}
\centering
\end{figure}

\begin{figure*}
  \centering
  \begin{tabular}{cc}
    \includegraphics[width=0.45\textwidth]{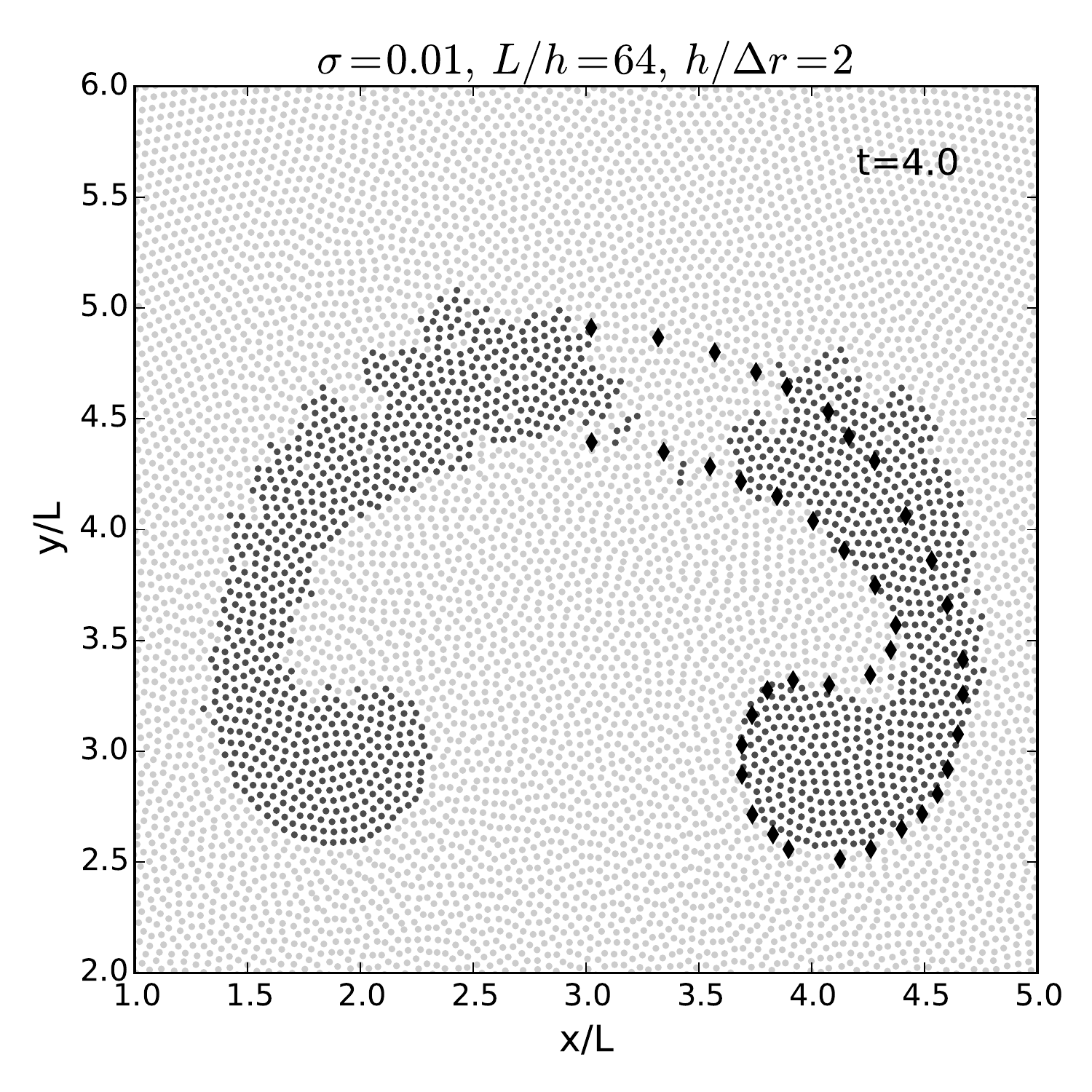} &
    \includegraphics[width=0.45\textwidth]{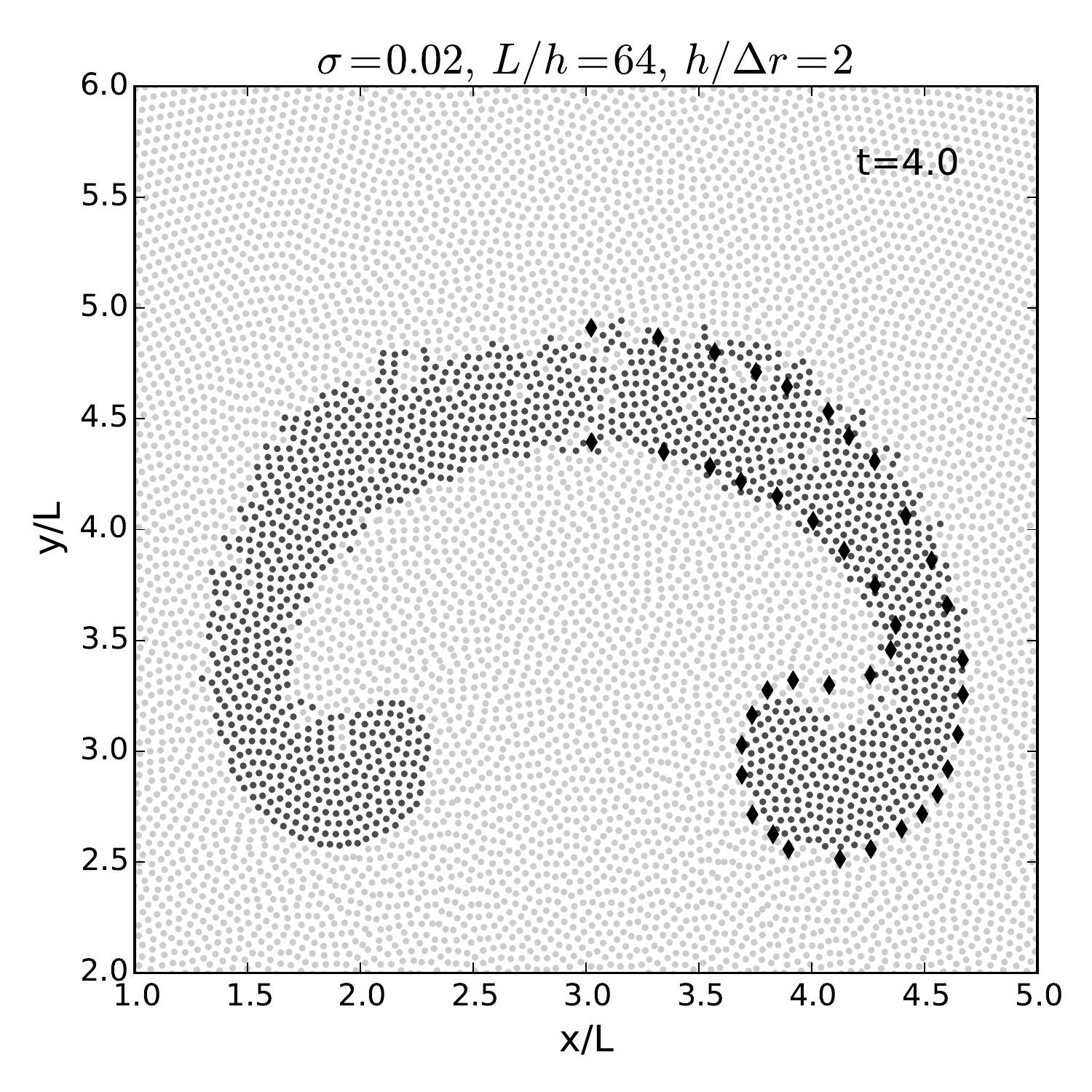} \\
  \end{tabular}
  \caption{Two-dimensional bubbles rising in liquid at $t=4$; the results obtained for $\sigma=0.01$ and $\sigma=0.02$. }
  \label{fig:bubbles-st}
\end{figure*}

Figures \ref{fig:bubbles-st} presents the particle positions (for $h=0.09375$ and $k=16.75$) at $t=4$ calculated for two different values of surface tension coefficient: $\sigma=0.01$ and $0.02$. The instability is clearly visible for $\sigma=0.01$. 
\begin{figure*}
  \centering
  \begin{tabular}{ccc}
    \includegraphics[width=0.31\textwidth]{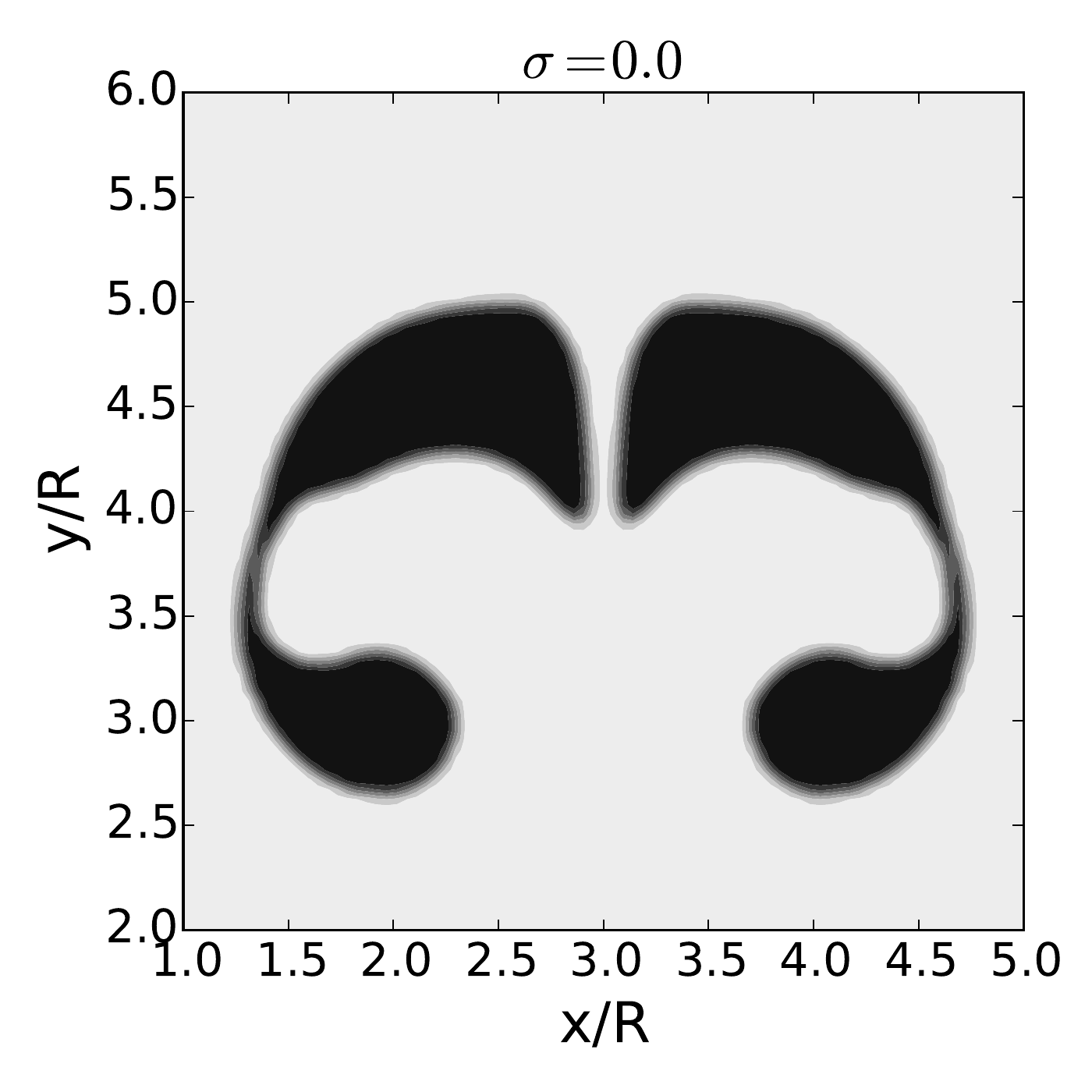} &
    \includegraphics[width=0.31\textwidth]{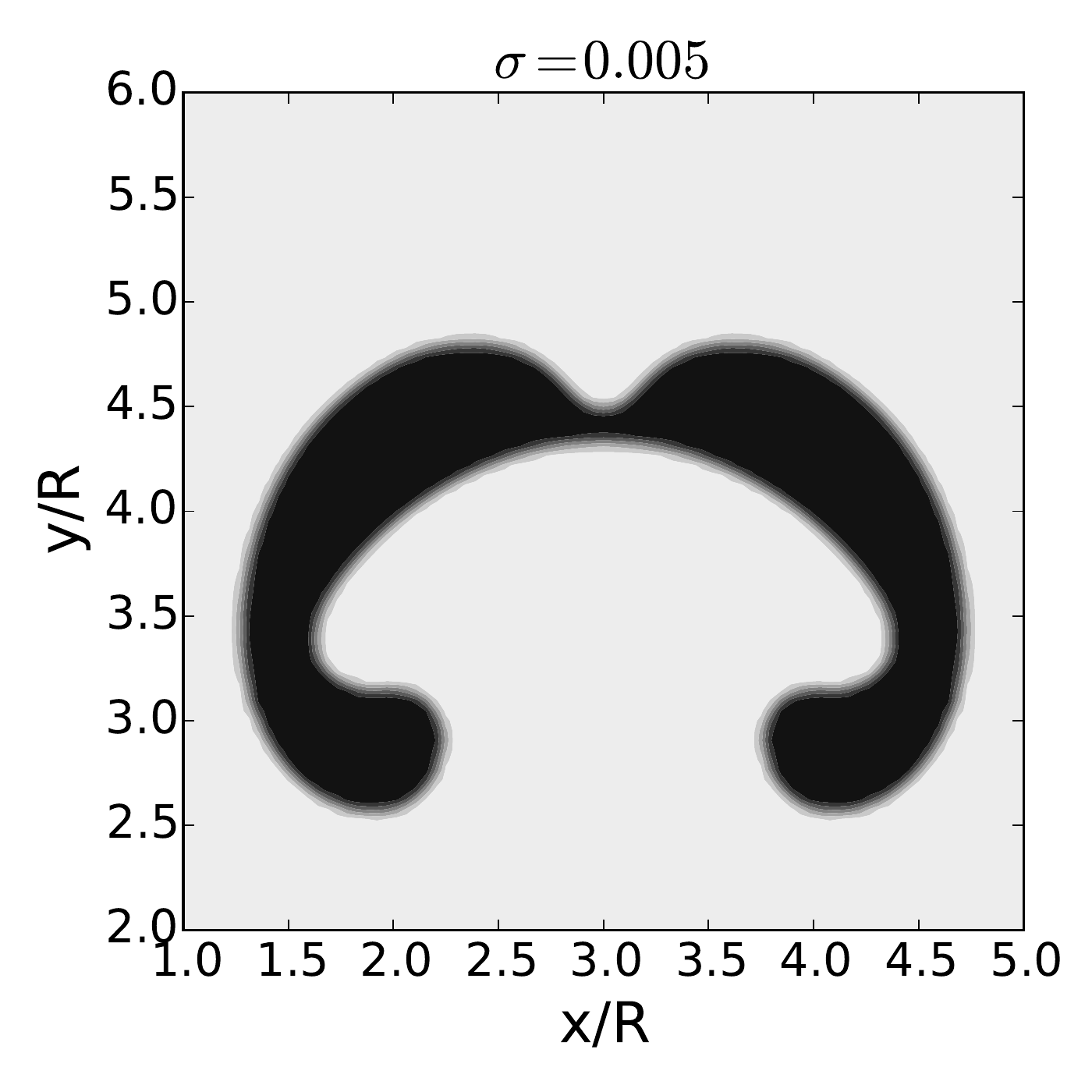} &
    \includegraphics[width=0.31\textwidth]{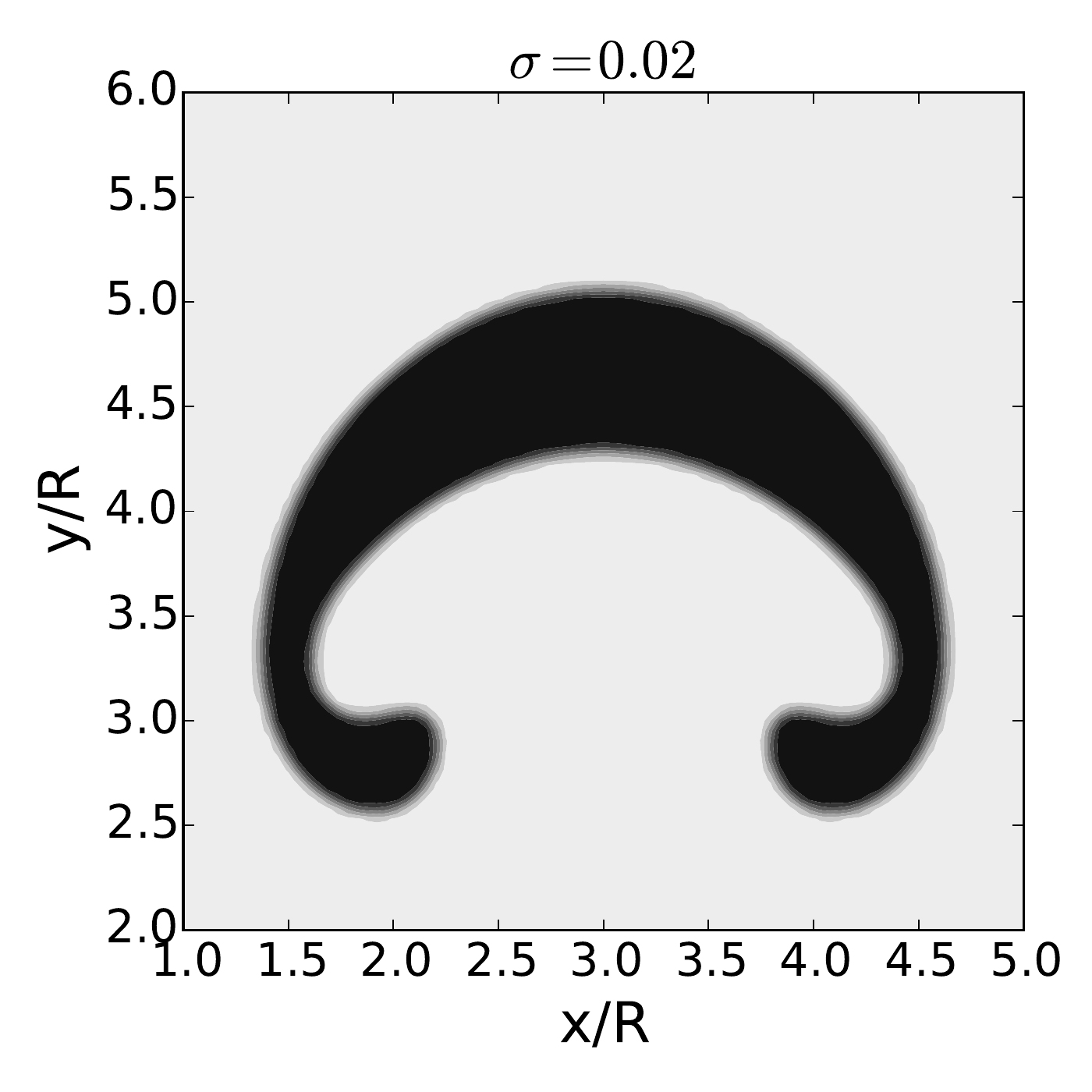}\\
  \end{tabular}
  \caption{The bubble rising in liquid test case at $t=4$ calculated using the VOF approach; the results obtained for $\sigma=0.0$, $0.005$ and $0.02$; the domain was composed of $128\times 256$ cells.}
  \label{fig:bubbles-vof}
\end{figure*}

In order to confirm the obtained results, we decided to perform calculations of the same test case using the Eulerian approach. The simulations were carried out with the open source CFD software \emph{OpenFOAM} (Finite Volume Method). We used the incompressible two-phase solver \emph{interFoam}, where the interface handling is solved using the Volume-Of-Fluid approach. The domain was composed of $128\times 256$ cells.
The simulations were performed for three different surface-tension coefficients: $\sigma=0.0$, $0.005$ and $0.02$. The obtained results at $t=4$ are presented in Fig.~\ref{fig:bubbles-vof}. The instability is clearly visible for $\sigma < 0.02$.

\begin{figure*}
  \centering
  \begin{tabular}{cc}
    \includegraphics[width=0.45\textwidth]{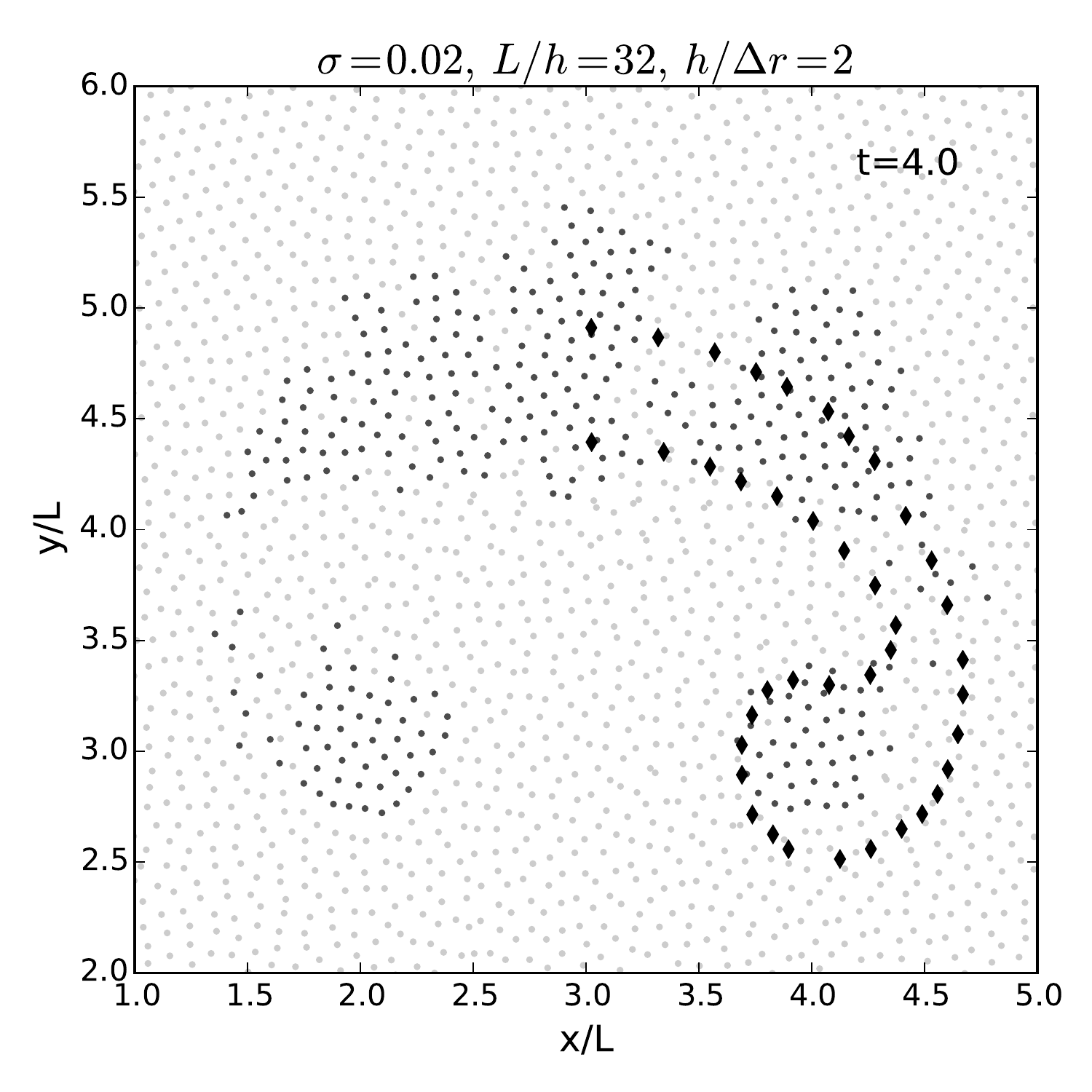} &
    \includegraphics[width=0.45\textwidth]{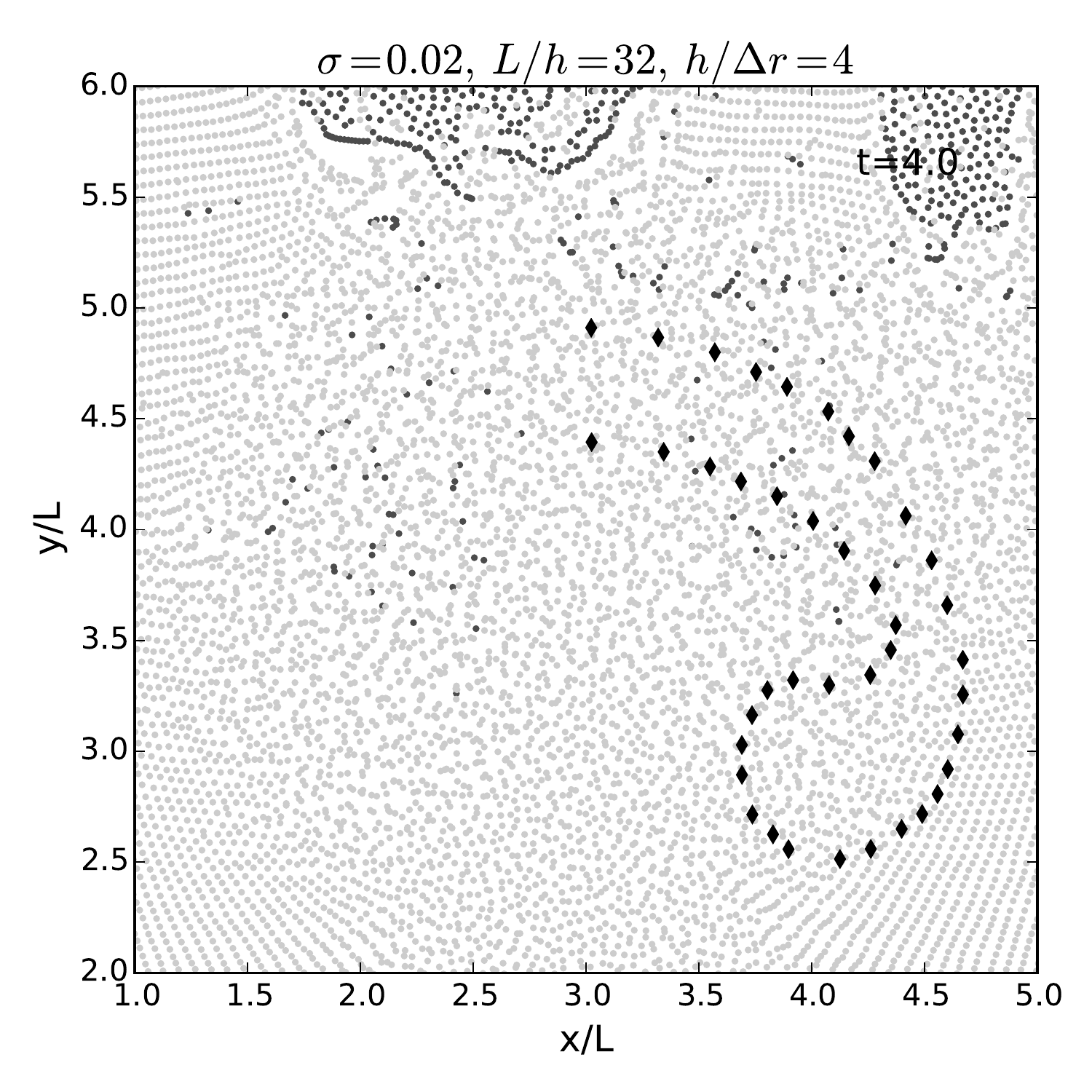} \\
  \end{tabular}
  \caption{Two-dimensional bubble rising in liquid at $t=4$; the results obtained for $\sigma=0.02$; (left) $h\Delta r=2$, $L/h=64$, $k=8.375$, (right)$h\Delta r=4$, $L/h=64$, $k=8.375$.}
  \label{fig:bubbles-st-varr}
\end{figure*}

\begin{figure*}
  \centering
  \begin{tabular}{cc}
    \includegraphics[width=0.45\textwidth]{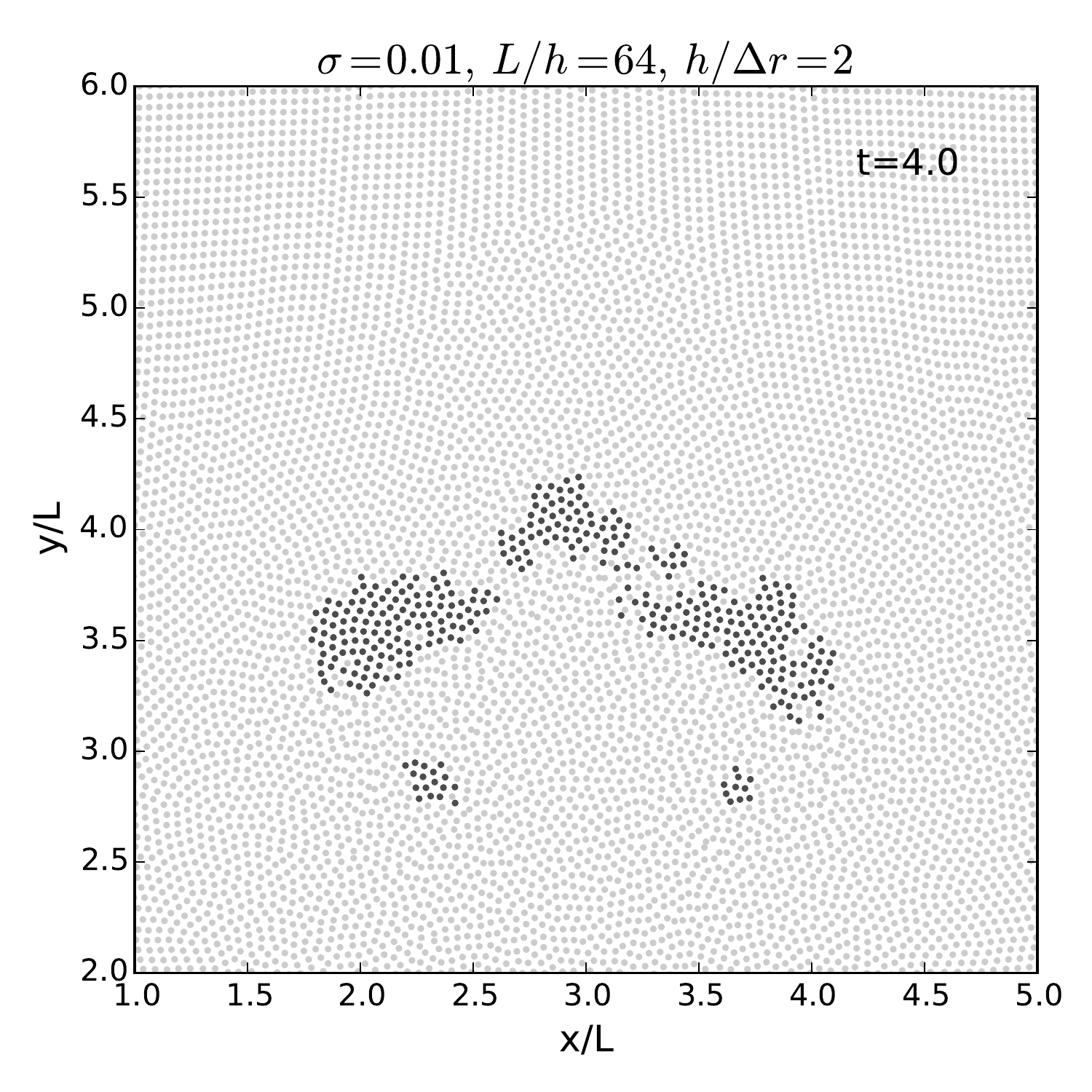} &
    \includegraphics[width=0.45\textwidth]{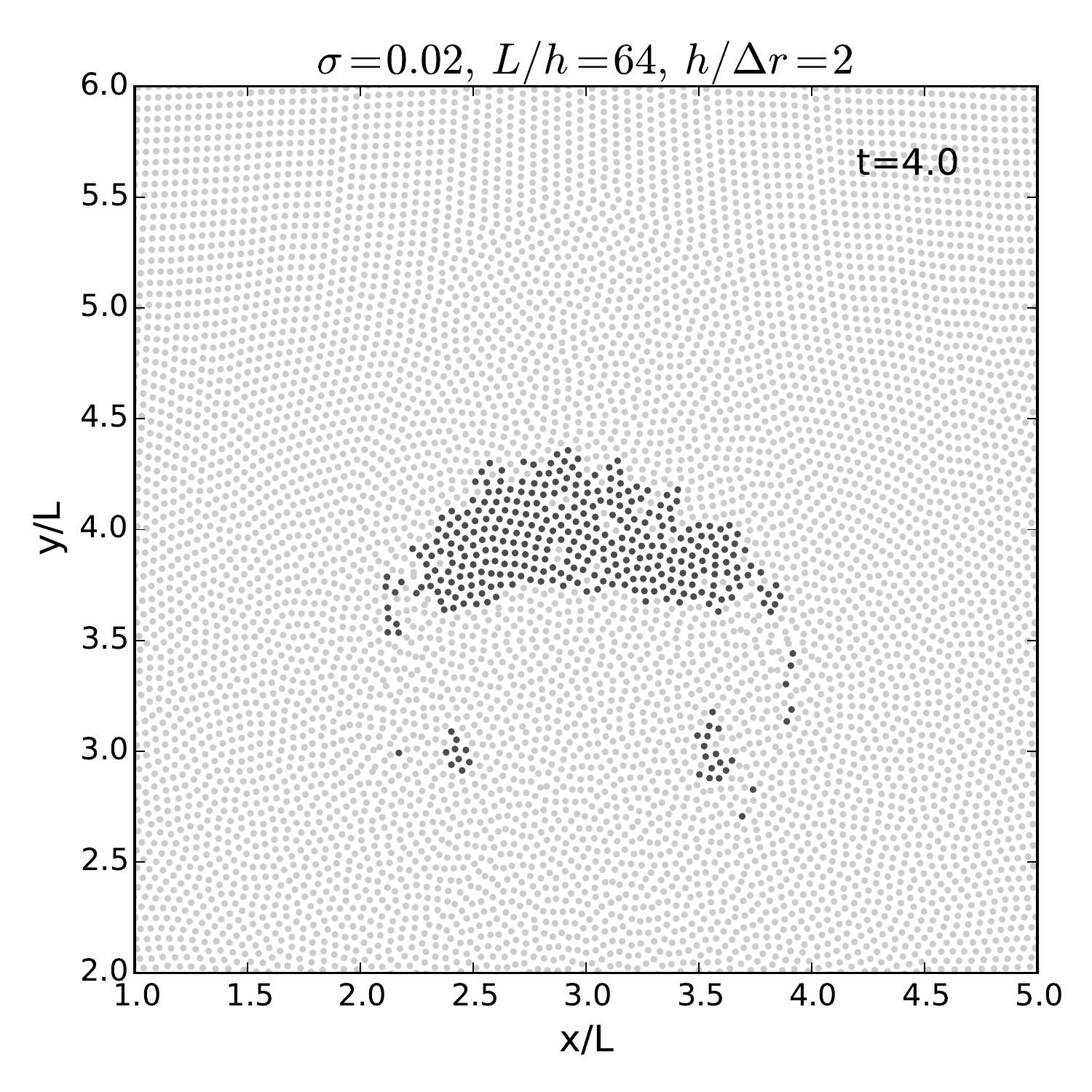} \\
  \end{tabular}
  \caption{Two-dimensional bubble rising in liquid at $t=4$; the results obtained for $\sigma=0.01$ and $\sigma=0.02$; the initial radius of bubble was $R/2$.}
  \label{fig:bubbles-st-vard}
\end{figure*}

In order to proof that the wavelength number of the disturbance is associated with $h$ and not with $\Delta r$ (which is of the same order of magnitude in the considered test case), we decided to perform the same test case as previously (stable solution, $\sigma=0.02$) but with lower $h=0.1875$ and two values of $h/\Delta r = 2$ and $4$. The first solution is presented in Fig.~\ref{fig:bubbles-st-varr}. It involves the same $h/\Delta r$ and lower $L/h$ than in the stable solution presented in Fig.~\ref{fig:bubbles-st}(right).
The second simulation involves twice higher $h/\Delta r$ and half $L/h$ than in the stable solution presented in Fig.~\ref{fig:bubbles-st}(right). It gives the same number of particles as in this reference test case. The obtained calculation is presented in Fig. \ref{fig:bubbles-st-varr}(right). Both results show a strong instability, which proves that $k$ is $h$-dependent. 
The obtained results we also confirmed for a small initial radius of bubble, $R/2$, see Fig.~\ref{fig:bubbles-st-vard}. 

\begin{figure*}
  \centering
  \includegraphics[width=0.7\textwidth]{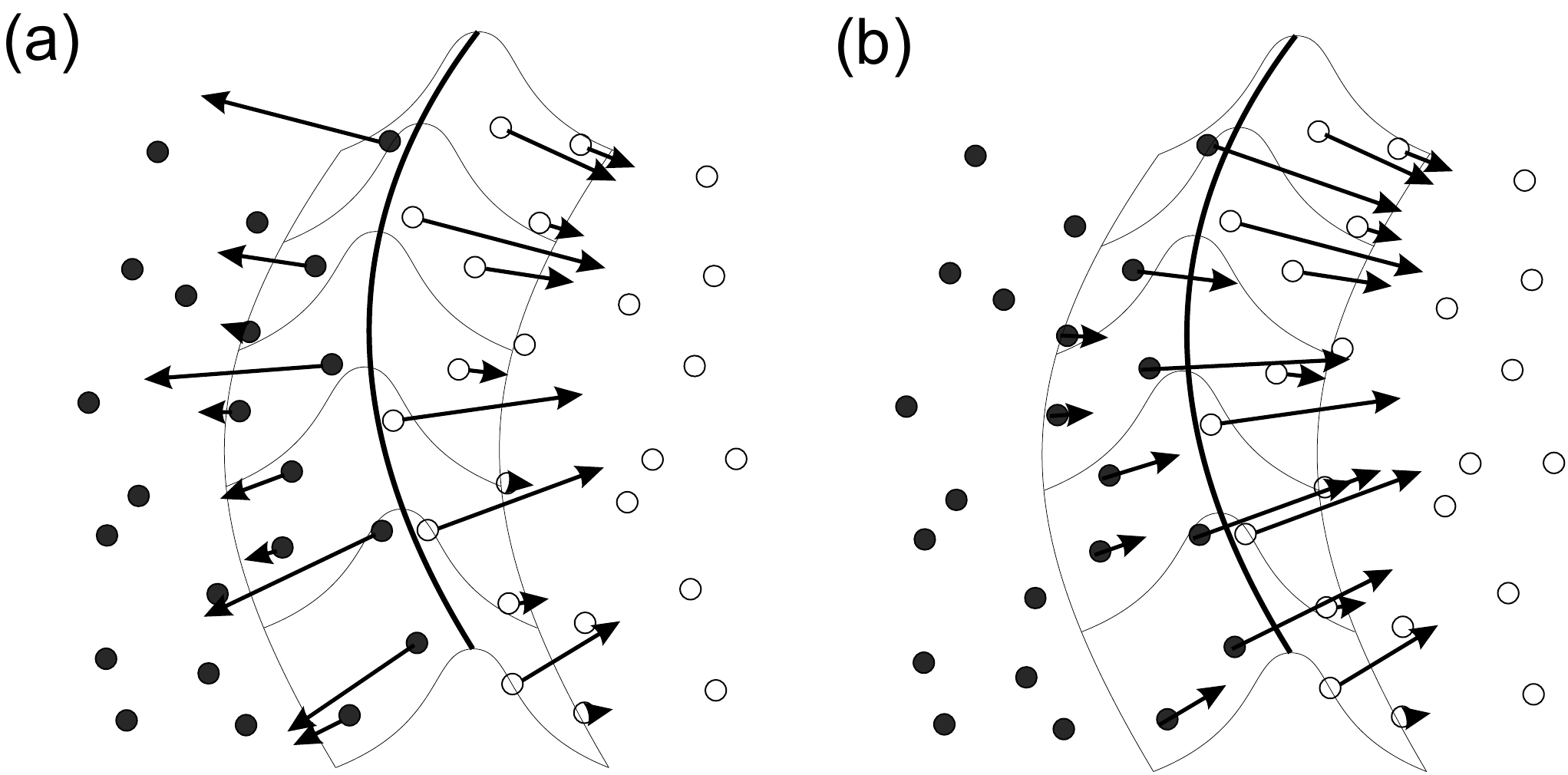}
  \caption{Sketch of the transition zone at the interface; (a) the repulsion provided by the interface sharpness correction procedure, (b) the interface surface tension force provided by the Continuum Surface Force technique, see \cite{Brackbill et al. 1992}.}
  \label{fig:corint-csf}
\end{figure*}

\begin{figure*}
  \centering
  \begin{tabular}{cc}
    \includegraphics[width=0.45\textwidth]{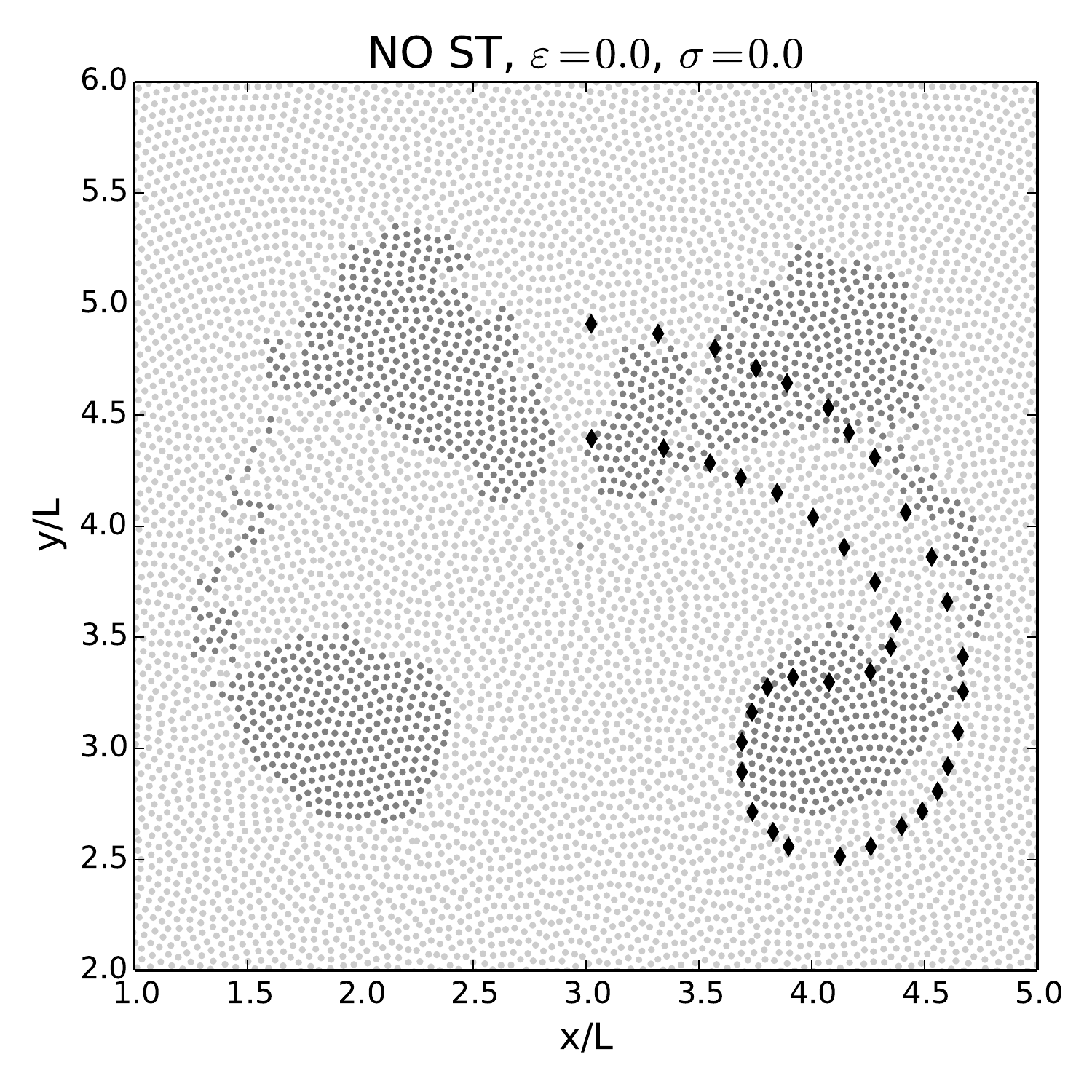} &
    \includegraphics[width=0.45\textwidth]{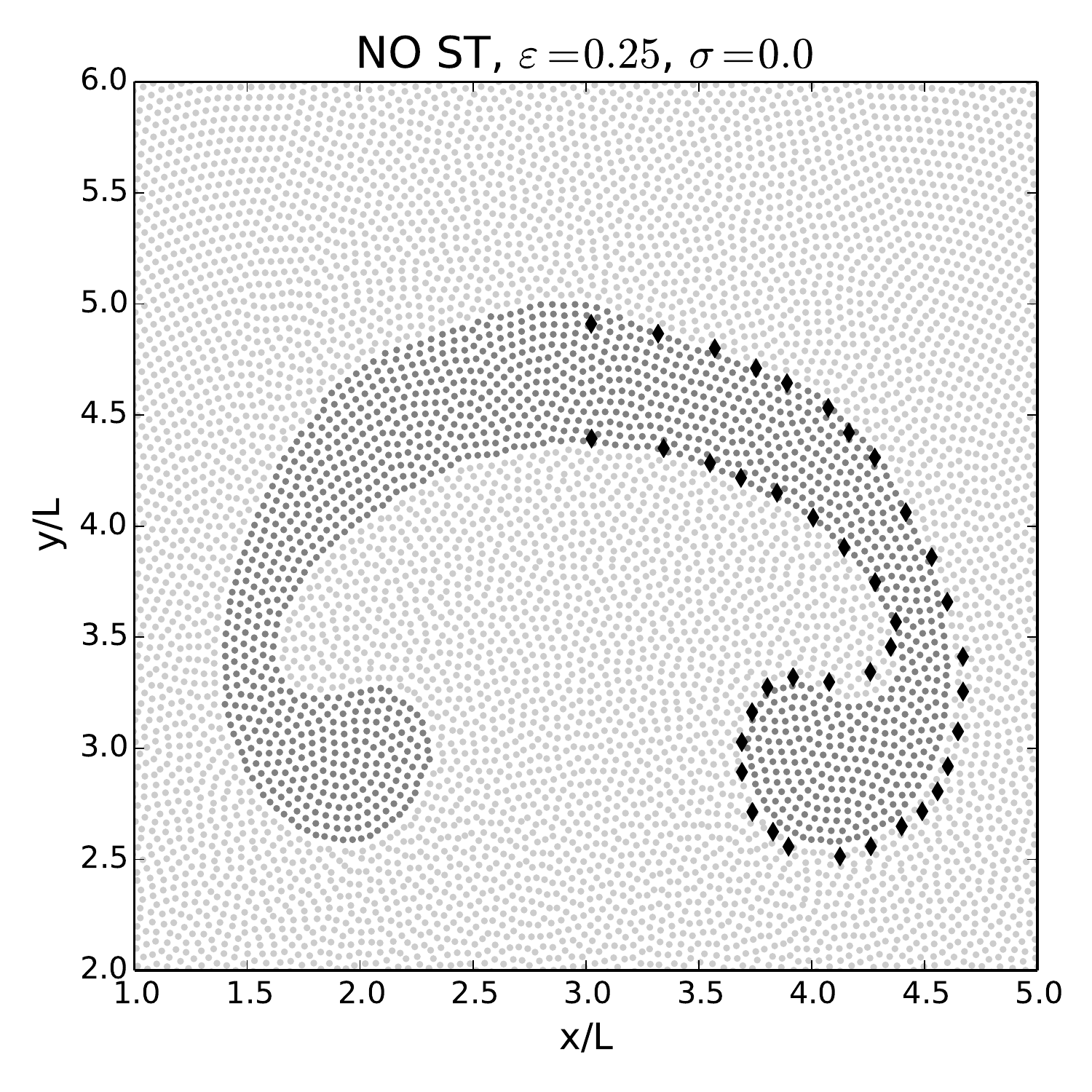} \\
  \end{tabular}
  \caption{Two-dimensional bubble rising in liquid at $t=4$ ($Re=1000$); particle positions obtained for $h/\Delta r = 2$, $L/h = 64$ and no surface tension; (left) the interface sharpness correction is not present, (right) $\varepsilon=0.25$.}
  \label{fig:bubbles_1}
\end{figure*}

We have shown that the surface tension can stabilize the interface so that the Kelvin-Helmholtz instability does not appear. Therefore, we conclude that the described in \cite{Szewc et al. 2015, Colagrossi and Landrini 2003, Das and Das 2009, Grenier et al. 2009} the non-physical behavior of bubbles, is actually physically correct. In the paper \cite{Szewc et al. 2015}, in order to remove the appearing instabilities, it was decided to apply the interface sharpness correction procedure by adding to the Navier-Stokes equation the new term in the form (\ref{xi}). This procedure, similarly to the surface tension in Continuum Surface Force model~\cite{Brackbill et al. 1992}, acts only at the interface (within the smoothing range), see Fig.~\ref{fig:corint-csf}. Due to the way how it is implemented, see \cite{Szewc et al. 2015} for details, similarly to the surface tension it stabilizes the interface preventing from the development of the Kelvin-Helmholtz instability. The influence of the interface sharpness correction on the test case considered in the previous sections is presented in Fig. \ref{fig:bubbles_1}.

However, even if a high surface tension is present and the system is stable, some of the particles belonging to different phases cross the interface, e.g. see Fig.~15 in \cite{Szewc et al. 2015}. In our opinion this is the Lagrangian manifestation of the numerical interface diffusion -- the issue known, under slightly altered form, in the Eulerian methods such as VOF. 

\begin{figure*}
  \centering
  \begin{tabular}{cc}
    \includegraphics[width=0.45\textwidth]{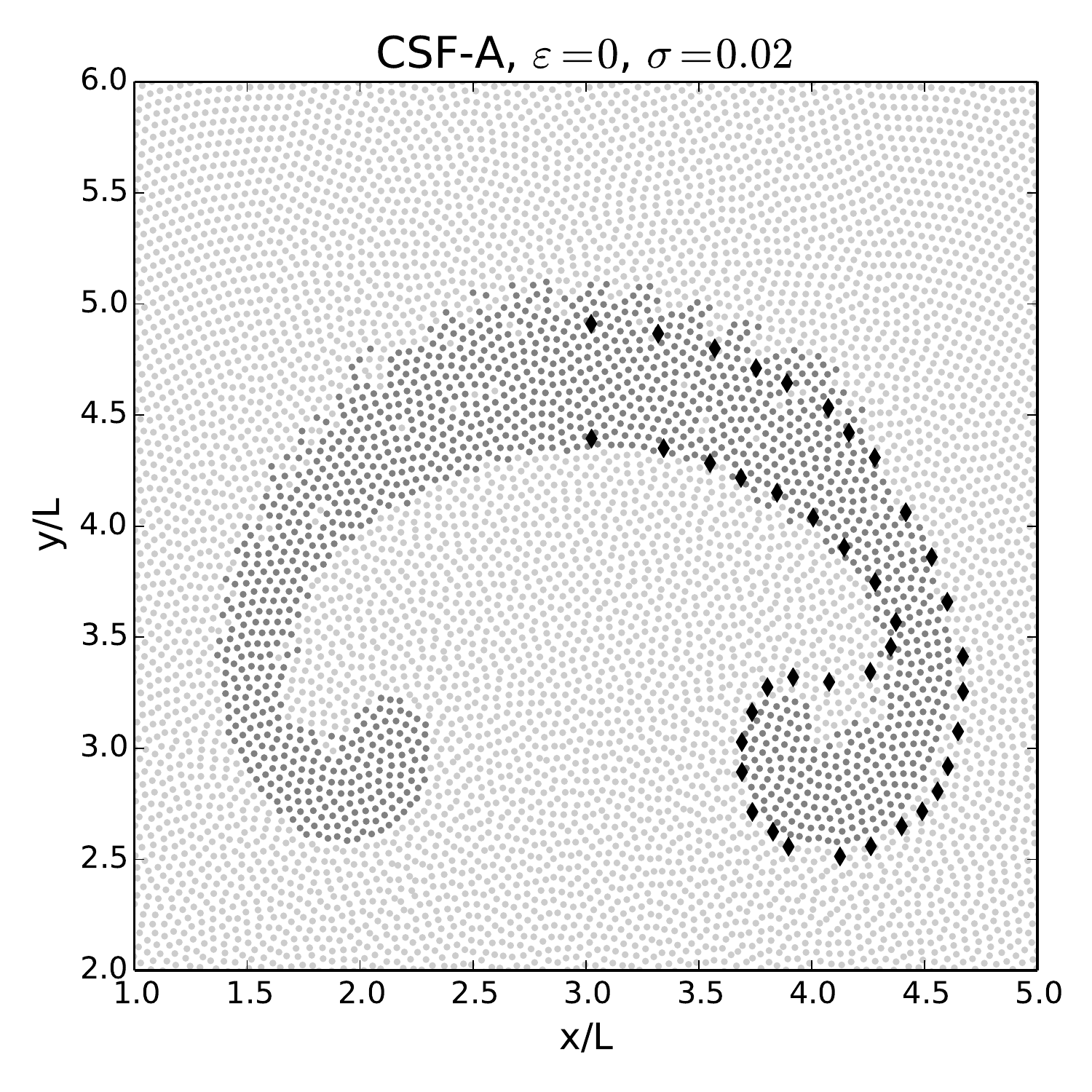} &
    \includegraphics[width=0.45\textwidth]{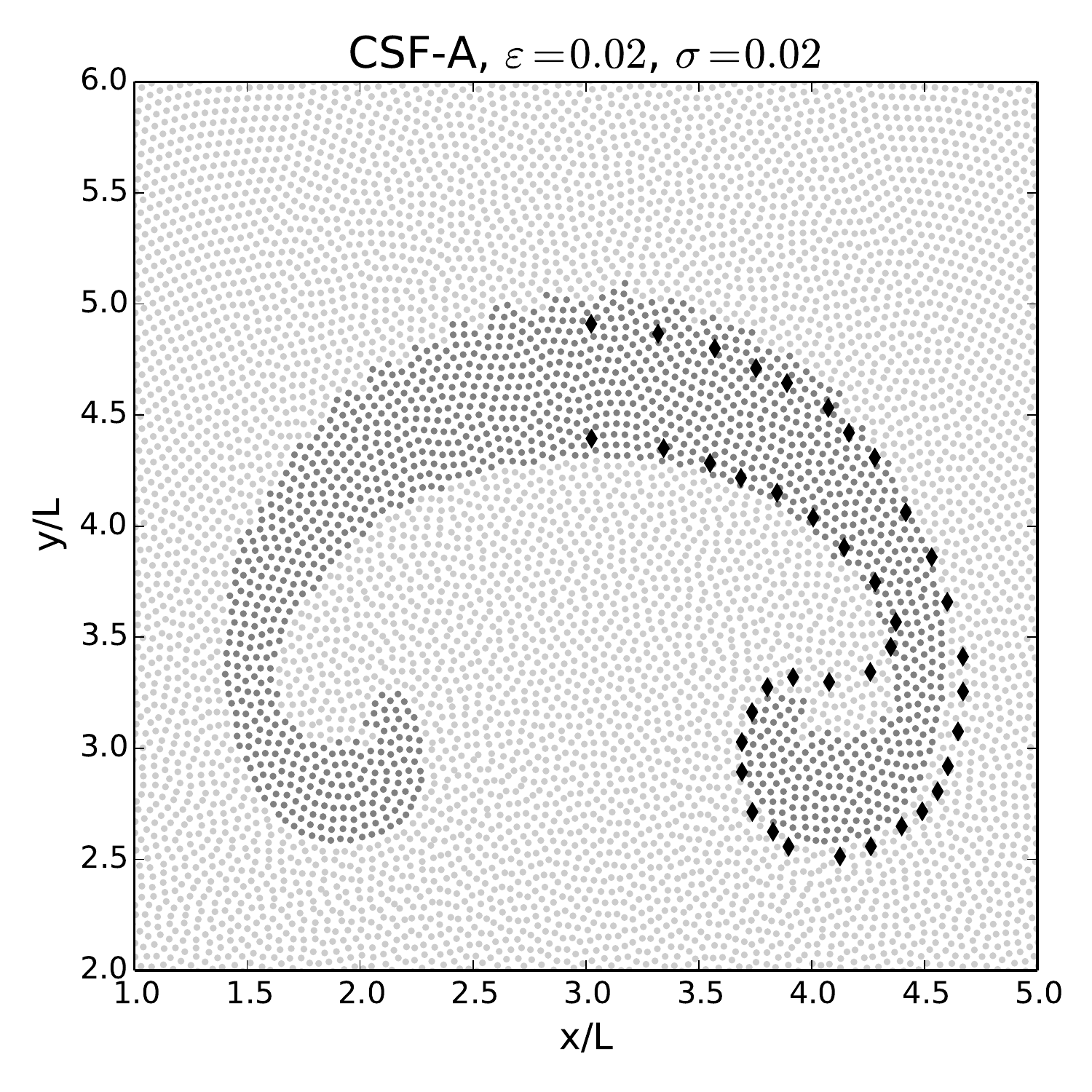} \\
    \includegraphics[width=0.45\textwidth]{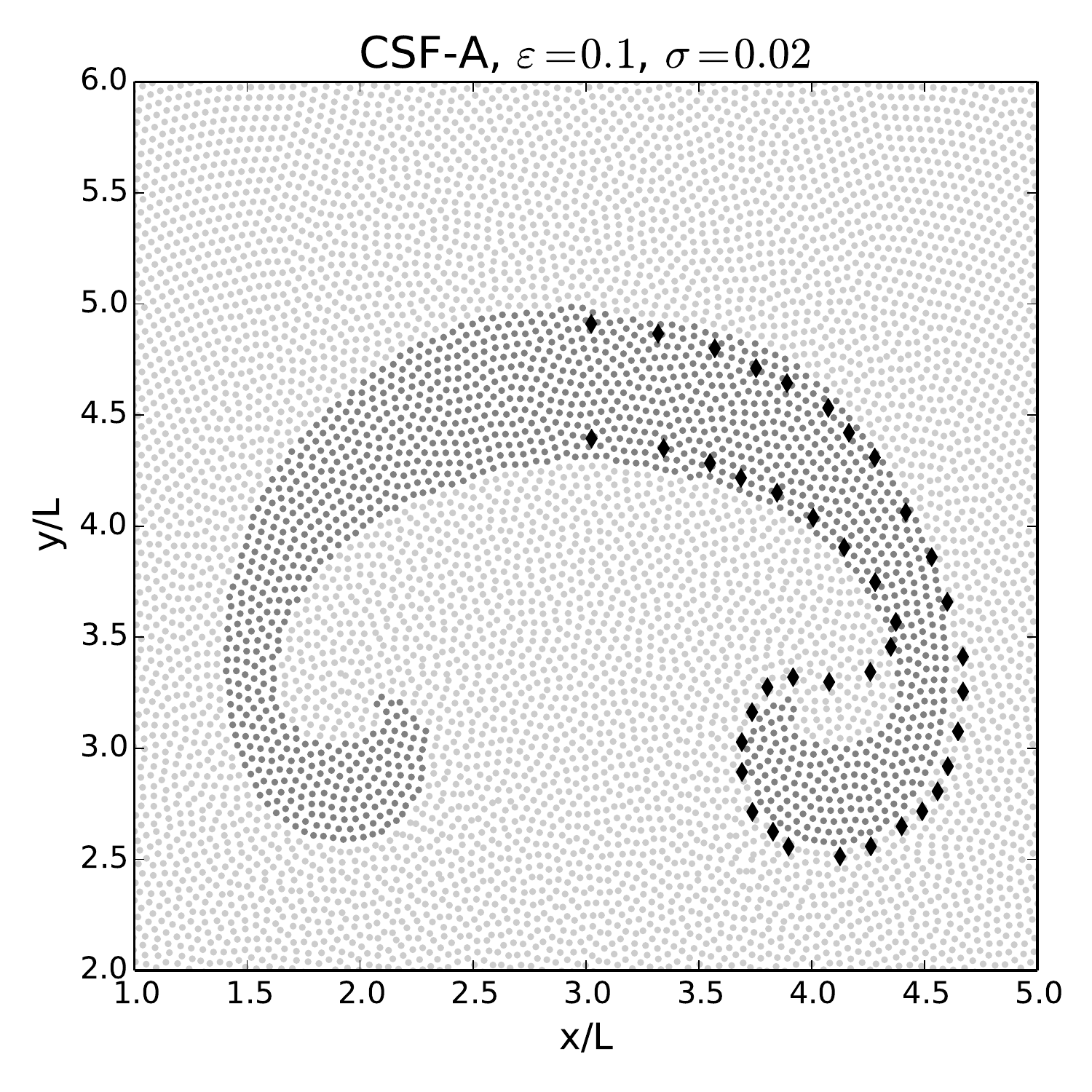} &
    \includegraphics[width=0.45\textwidth]{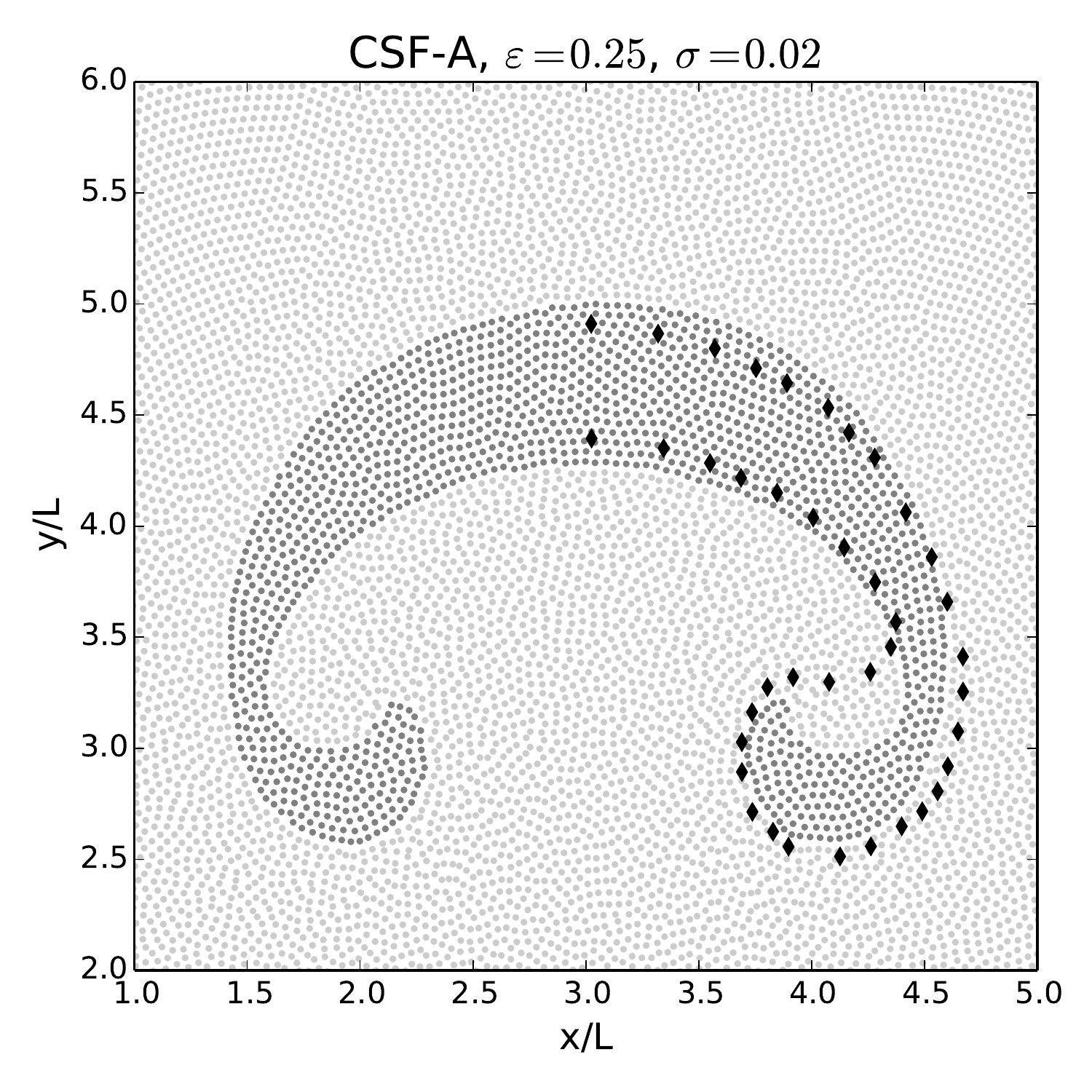} \\
  \end{tabular}
  \caption{Two-dimensional bubble rising in liquid at $t=4$ ($Re=1000$); particle positions obtained for $h/\Delta r = 2$, $L/h = 64$; the surface tension is present ($\sigma=0.02$); results obtained for different values of the surface sharpness correction coefficient: $\varepsilon=0$, $0.02$, $0.1$ and $0.25$.}
  \label{fig:bubbles_2}
\end{figure*}

Bearing in mind the undertaken stability analysis, it is important to note that for the SPH modeling of processes involving the surface tension, the interface sharpness correction term seems to have a beautifying role only, see Fig.~\ref{fig:bubbles_2}.

\section{The interface sharpness correction term, relation between $\varepsilon$ and $h$} \label{sec:e-h}
In the paper \cite{Szewc et al. 2015} the authors studied the dependency between the minimum interface sharpness correction coefficient $\varepsilon_{min}$ and the smoothed length $h$. The fully numerical analysis gave us the relation
\begin{equation} \label{micromixing numerical dependency}
  \varepsilon_{min} \sim \frac{1}{h}.
\end{equation}
This solution was obtained for the same test case of the two-dimensional bubble rising in liquid as in the previous sections. In this section, we show that this result can be predicted from the simple stability analysis.
Let us start with replacing Eq.~(\ref{pressure jump with surface tension}) with the normal stress condition in the form 
\begin{equation} \label{pressure jump with interface correction}
  p_U - p_L = \varepsilon.
\end{equation} 
Analogously to the analysis performed in the previous sections, the pressure constraint (\ref{pressure jump with interface correction}) requires that
\begin{equation}
  \varrho_L\left[gk - (s+iku_L)^2 \right] - \varrho_U\left[gk + (s+iku_U)^2 \right] = \frac{k}{\xi_0} \varepsilon.
\end{equation}
Solving this quadratic equation we obtain
\begin{equation}
  s = -ik \frac{\varrho_U u_U + \varrho_L u_L}{\varrho_U + \varrho_L} \pm \left[ \frac{k^2 \varrho_U \varrho_L (u_U - u_L)^2}{(\varrho_U + \varrho_L)^2} + kg \frac{\varrho_U - \varrho_L}{\varrho_U + \varrho_L} - \frac{k}{\xi_0} \frac{\varepsilon}{\varrho_U + \varrho_L} \right]^{\frac{1}{2}}.
\end{equation}
This system is stable if and only if the quantity in square brackets is negative
\begin{equation} \label{micromixing condition}
  \frac{k^2 \varrho_U \varrho_L (u_U - u_L)^2}{(\varrho_U + \varrho_L)^2} + kg \frac{\varrho_U - \varrho_L}{\varrho_U + \varrho_L} - \frac{k}{\xi_0} \frac{\varepsilon}{\varrho_U + \varrho_L} < 0.
\end{equation}
Assuming that $k = 2\pi / 4h$, Eq.~(\ref{micromixing condition}) takes the form
\begin{equation}
  \varepsilon > \frac{A}{h} + B,
\end{equation}
where
\begin{equation}
A=\frac{\pi \varrho_U \varrho_L (u_U-u_L)^2 \xi_0}{2(\varrho_U + \varrho_L)},\quad B = g(\varrho_U - \varrho_L) \xi_0.
\end{equation}
This result agrees with the numerically obtained dependency (\ref{micromixing numerical dependency}).

\section{Conclusions}
On the basis of both: the analytic stability analysis and the numerical calculations using SPH and VOF approaches, we have shown that the surface tension can stabilize the interface so that the Kelvin-Helmholtz instability does not appear. In the test case introduced by \emph{Sussman et al. (1994)}~\cite{Sussman et al. 1994}, the surface tension force appeared to be negligible compared to other forces, however, it is strong enough to be responsible for stabilization of the interface. Many researchers, in order to validate the SPH models, decided to compare their solution with this case, but, with the surface tension effects neglected, see~\cite{Szewc et al. 2015, Colagrossi and Landrini 2003, Grenier et al. 2009}. To remove the appearing instabilities, these authors decided to introduce the interface correction procedures. In this work we have demonstrated that these procedures may lead to non-physical solutions. Therefore, since the interface correction procedure seems to have a beautifying role only, we recommend to use the interface correction procedure with caution. In this paper, on the basis on the stability analysis, we have also explained the puzzling relation between the parameter $\varepsilon$ and $h$ introduced in \cite{Szewc et al. 2015}.

\section*{Acknowledgments}
This research has been partly funded by the \'Electricit\'e de France R\&D, Chatou, France.


\begin{thebibliography}{9}


\bibitem{Colagrossi and Landrini 2003}
Colagrossi~A, Landrini~M. 
Numerical simulation of interfacial flows by smoothed particle hydrodynamics,
\emph{Journal of Computational Physics} 2003; \textbf{191} : 227-264.

\bibitem{Das and Das 2009}
Das~AK, Das~PK.
Bubble evolution through submerged orifice using smoothed particle hydrodynamics:
Basic formulation and model validation.
\emph{Chemical Engineering Science} 2009; \textbf{64} : 2281-2290.

\bibitem{Grenier et al. 2009}
Grenier~N, Antuono~M, Colagrossi~A, Le Touz\'e~D, Alessandrini~B. 
An Hamiltonian interface SPH formulation for multi-fluid and free-surface flows. 
\emph{Journal of Computational Physics} 2009; \textbf{228} : 8380-8393.

\bibitem{Szewc et al. 2015}
Szewc~K, Pozorski~J, Minier~JP.
Spurious interface fragmentation in multiphase SPH.
\emph{International Journal for Numerical Methods in Engineering} 2015; \textbf{103} : 625-649.

\bibitem{Sussman et al. 1994}
Sussman~M, Smereka~P, Osher~SJ.
A level-set approach for computing solutions to incompressible two-phase flow.
\emph{Journal of Computational Physics} 1994; \textbf{114} : 146-159.

\bibitem{Mestel}
Mestel~AJ. 
Hydrodynamic Stability.
\emph{http://wwwf.imperial.ac.uk/{\textasciitilde}ajm8/}.


\bibitem{Brackbill et al. 1992}
Brackbill~JU, Kothe~DB, Zemach~C.
A continuum method for modelling surface tension.
\emph{Journal of Computational Physics} 1992; \textbf{100} : 335-354.


\end{thebibliography}
\end{document}